\documentclass[
11pt,%
final,%
tightenlines,%
notitlepage,%
twoside,%
onecolumn,%
nobibnotes,%
nofootinbib,%
superscriptaddress,%
noshowpacs,%
noshowydk,%
showdoi,%
centertags,%
maik]%
{revtex4n}

\input jpack_e
\setcaptionmargin{5mm}
\onelinecaptionstrue

\columntovsizetrue
\allowdisplaybreaks
\begin{document}

\newtheorem{teo}{Theorem}
\newtheorem*{teon}{Theorem}
\newtheorem{lem}{Lemma}[section]
\newtheorem*{lemn}{Lemma}
\newtheorem{prp}{Proposition}[section]
\newtheorem*{prpn}{Proposition}
\newtheorem{ass}{Assertion}
\newtheorem*{assn}{Assertion}
\newtheorem{assum}{Assumption}
\newtheorem*{assumn}{Assumption}
\newtheorem{stat}{Statement}
\newtheorem*{statn}{Statement}
\newtheorem{cor}{Corollary}
\newtheorem*{corn}{Corollary}
\newtheorem{hyp}{Hypothesis}
\newtheorem*{hypn}{Hypothesis}
\newtheorem{con}{Conjecture}
\newtheorem*{conn}{Conjecture}
\newtheorem{dfn}{Definition}[section]
\newtheorem*{dfnn}{Definition}

\newtheorem{problem}{Problem}
\newtheorem*{problemn}{Problem}
\newtheorem{notat}{Notation}
\newtheorem*{notatn}{Notation}
\newtheorem{quest}{Question}
\newtheorem*{questn}{Question}

\theorembodyfont{\rm}
\newtheorem{rem}{Remark}
\newtheorem*{remn}{Remark}
\newtheorem{exa}{Example}
\newtheorem*{exan}{Example}
\newtheorem{cas}{Case}
\newtheorem*{casn}{Case}
\newtheorem{claim}{Claim}
\newtheorem*{claimn}{Claim}
\newtheorem{com}{Comment}
\newtheorem*{comn}{Comment}

\theoremheaderfont{\it}
\theorembodyfont{\rm}

\newtheorem{proof}{Proof}[section]
\newtheorem*{proofn}{Proof}

\theoremstyle{plain}


\selectlanguage{english}
\Rubrika{\relax}
\CRubrika{\relax}
\SubRubrika{\relax}
\CSubRubrika{\relax}

\newcommand{\zt}[0]{\Tilde{z}}
\newcommand{\psit}[0]{\Tilde{\psi}}
\newcommand{\rt}[0]{\Tilde{r}}
\newcommand{\zb}[0]{\bar{z}}
\newcommand{\rb}[0]{\bar{r}}
\newcommand{\Zb}[0]{\bar{Z}}
\newcommand{\Rb}[0]{\bar{R}}
\newcommand{\econe}{\epsilon_{c_1}}
\newcommand{\ectwo}{\epsilon_{c_2}}
\newcommand{\ec}{\epsilon_{c}}
\newcommand{\Uinfd}{U_{\infty}^{\ast}} 
\newcommand{\Uinf}{U_{\infty}} 
\newcommand{\Gd}{G^{\ast}_{\infty}} 
\newcommand{\psid}{\psi^{\ast}_{\infty}} 
\newcommand{\dy}{\mathrm{d}\mathbf{y}}
\newcommand{\ints}{\mathbb{Z}}
\newcommand{\psiinf}{\psi_{\infty}}
\newcommand{\psiinft}{\tilde{\psi}_{\infty}}
\newcommand{\Ginf}{G_{\infty}}

\newcommand{\Lf}{\Lambda_f}

\newcommand{\karman}{von K\'arm\'an}
\newcommand{\benard}{B\'enard}

\newcommand{\connected}{connected}

\newcommand{\bigo}{\mathcal{O}}
\renewcommand{\epsilon}{\varepsilon}


\def\JournalNumber{0}
\def\JournalVolume{00}
%
%
%
\nameVolumeRus{}
\CnameVolumeRus{}
\nameIssueRus{\No}
\CnameIssueRus{}
\namePartRus{}
\namePagesRus{}
\nameYearShortRus{}
\JournalNameRus{}
\TranslitJournalNameRus{}
\JournalName{Regular and Chaotic Dynamics}
\JournalISSNCode{1560-3547}
\IssuePrice{}
\TransYearOfIssue{0000}
\TransCopyrightYear{2016}%
\OrigYearOfIssue{}
\OrigCopyrightYear{2016}%
\OrigIssueNo{\JournalNumber}
\OrigVolumeNo{\JournalVolume}
\TransVolumeNo{\JournalVolume}
\TransIssueNo{\JournalNumber}
\TransPartNo{}
\SHORTjournalPREFIX{RCD} 
\LONGjournalPREFIX{RegDyn} 
\BatFileName{call make_ps.bat} 
\BatSwitch{3} 
\IssueName{}
\SupplementNumber{}
\PublicationSerialNumberInYear{0}
\PublicationSerialNumberInVolume{0}
\ConditionalIssueDate{"year","month","day","name","type"}
\PagePrefix{}
\JournalISSNonlineCode{}
\JournalISSNCodeRus{}
\JournalISSNonlineCodeRus{}
\VolumeName{}
\IssnoName{none}
\PartnoName{}
\FpageNamepp{}
\FpageNnamep{}
\FpagePrefix{}
\LpageNnamepp{}
\LpageNamep{}
\LpagePrefix{}
\VolumePageNumbering{}
\JournalPubID{}
\FirstJournalPageNumber{}
\LastJournalPageNumber{}
\makeatletter
\def\MAIKlogo{RCD Editorial Office}
\def\maikpraefix{10.0000/S}
\edef\@ContentsHeadLineB{Simultaneous English language translation of the journal is available from \noexpand\MAIKlogo}
\def\Distributed{Distributed worldwide by Springer. }
\def\ArticlePages#1{\relax}
\@ifxundefined\CONT@sw{\@booleantrue\CONT@sw}{}%
\@booleantrue\showPACS@sw%
\@booleantrue\showKEYS@sw %
\@booleantrue\noOrigJournalVersion@sw
\@booleantrue\noOrigVolumeNo@sw
\@booleanfalse\noTransVolumeNo@sw
\makeatother
\input maikdoi %

\beginpaper


\input engnames
\titlerunning{Atmosphere topology of vortex ring arrays}
\authorrunning{Masroor, Stremler}
\toctitle{Title}
\tocauthor{F.\,S.\,Author}
\title{On the topology of the atmosphere advected by a periodic array of axisymmetric thin-cored vortex rings}
\firstaffiliation{
}%
\articleinenglish 
\PublishedInRussianNo
\author{\firstname{Emad}~\surname{Masroor}}%
\email[E-mail: ]{emad@vt.edu}
\affiliation{
Engineering Mechanics Program\\
Virginia Polytechnic Institute \& State University\\
Blacksburg VA 24061, United States}%
\author{\firstname{Mark\,A.}~\surname{Stremler}}%
\email[E-mail: ]{stremler@vt.edu}
\affiliation{
Department of Biomedical Engineering \& Mechanics\\
Virginia Polytechnic Institute \& State University\\
Blacksburg VA 24061, United States}%
%
\begin{abstract}
The fluid motion produced by a periodic array of identical, axisymmetric, thin-cored  vortex rings is investigated. 
It is well known that such an array moves uniformly without change of shape or form in the direction of the central axis of symmetry, and is therefore an equilibrium solution of Euler's equations. In a frame of reference moving with the system of vortex rings, the motion of passive fluid particles is investigated as a function of the two non-dimensional parameters that define this system: $\epsilon = a/R$, the ratio of minor radius to major radius of the torus-shaped vortex rings, and $\lambda=L/R$, the separation of the vortex rings normalized by their radii. Two bifurcations in the streamline topology are found that depend significantly on $\epsilon$ and $\lambda$; these bifurcations delineate three distinct shapes of the `atmosphere' of fluid particles that move together with the vortex ring for all time.  Analogous to the case of an isolated vortex ring, the atmospheres can be `thin-bodied' or `thick-bodied'.  Additionally, we find the occurrence of a `\connected' system, in which the atmospheres of neighboring rings touch at an invariant ring of fluid particles that is stationary in a frame of reference moving with the rings.  
\end{abstract}
\received{December 11, 2021}
\revised{Month XX, 20XX}
\accepted{Month XX, 20XX}%
\maketitle

\textmakefnmark{0}{)}%

\section{Introduction}
Circular vortex rings were first identified as solutions of the Euler equations by von~Helmholtz~\cite{Helmholtz1858, tait1867pm}, who showed that a thin circular vortex filament will move uniformly along its central axis of symmetry without change of shape or form. While observations of vortex rings very likely preceded their mathematically rigorous definition in the 19th century --- Northrup~\cite{Northrup1911} points out that vortex rings in the form of `smoke rings' must surely have been recognized by people wherever tobacco smoking was practiced --- Helmholtz's paper led to a flurry of activity from the leading minds of nineteenth-century physics into the mathematical aspects of vortex rings,
with a significant driver of these early investigations being the vortex theory of atoms~\cite{Thomson1867a,Thomson1883}.

\begin{figure}
\centering
\includegraphics[width=155mm]{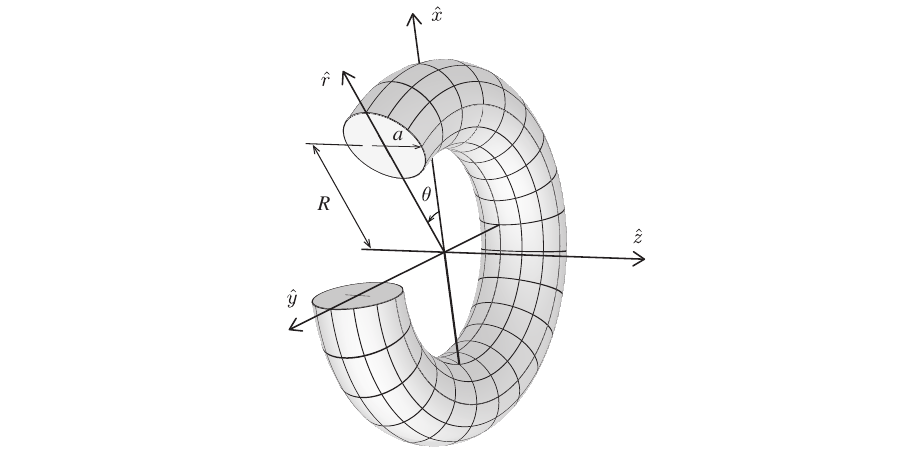}
\caption{
	An axisymmetric vortex ring in the cylindrical coordinate system $(\hat{z},\hat{r},\theta)$, with $\hat{z}$ the axial direction. 
	Cartesian coordinates $(\hat{x},\hat{y},\hat{z})$ are shown for reference. 
	The core radius is $a$ and the ring radius is $R$. 
	A $60^{\circ}$ arc of the ring has been removed from the schematic for illustration purposes.}
\label{fig:schematic}
\end{figure}

Following Helmholtz's seminal paper, a variety of vortex ring configurations have been investigated using analytical, experimental, and computational research methods. The greatest amount of attention has been paid to \emph{coaxial} sets of vortex rings, i.e., arrangements in which all of the vortex rings share a common central axis of symmetry and motion. Meleshko~\cite{Meleshko2010} compiled a bibliography of the literature on coaxial vortex rings, which ran to 238 entries by the year 2010. Because this system is axisymmetric, it is particularly amenable to a mathematical treatment in cylindrical coordinates $(\hat{z},\hat{r},{\theta})$, where $\hat{z}$ is the axial direction, $\hat{r}$ the radial direction, and ${\theta}$ the azimuthal direction, as illustrated in Figure~\ref{fig:schematic}; we will use a hat to denote dimensional variables.  In fact, the equations governing the motion of coaxial vortex rings can be cast in canonical form~\cite{Dyson1893,Vasilev1913,Borisov2013}, which has led to a large body of literature on coaxial vortex rings from the Hamiltonian perspective. 

Professor Alexei Borisov, to whom this memorial issue of \emph{RCD} is dedicated, made several important contributions to the study of vortex dynamics broadly and axisymmetric vortex rings specifically. 
Although the Hamiltonian nature of the equations of vortex rings has been known for a long time (implicitly in the works of Dyson~\cite{Dyson1893} and Hicks~\cite{Hicks1922}, and explicitly in the work of Vasilev~\cite{Vasilev1913}), Borisov et.~al~\cite{Borisov2013} were the first to systematically exploit the integrals of motion to express these equations in reduced canonical form for $N$ thin-cored vortex rings.

Coaxial sets of vortex rings can be observed in many physical and engineered fluid systems. 
In a body-centered frame of reference, steady flow past a sphere, a disk, or a ring~\cite{Bearman1988, Allen2007, Taylor1953, Sallet1975} can produce clear, coherent vortex rings.  
A standard method of producing repeatable coaxial vortex rings in the laboratory mimics the blowing of smoke rings: a piston is used to pulse a jet from a tube or through an orifice~\cite{Krueger2005,Krueger2009}.  
Vortex-producing pulsed jets have drawn interest as a means of propulsion for small underwater vehicles~\cite{Mohseni2006}.
The wakes of many marine animals can be viewed as a set of vortex rings being shed downstream of a swimming body~\cite{Gordon2017}. 
Coaxial vortex rings provide a mathematical framework for understanding the observed wakes of jellyfish in laboratory experiments~\cite{Dabiri2005}, and Siekmann~\cite{Siekmann1963} developed a formula for the thrust produced by an axisymmetric body whose wake takes the form of successive coaxial vortex rings, analogous to \karman's work on two-dimensional wakes. 
Ellington's \cite{Ellington1984} model of a hovering insect's wake consists of successive coaxial vortex rings.
Heinzel~\cite{Heinzel1987} has proposed pulsating vortex rings, generated by a periodic flick of the wings, as a possible method of communication between crickets.

Consider a toroidal-shaped region of fluid with non-zero vorticity embedded in an otherwise irrotational fluid that encompasses all space. 
We will refer to the coherent region of non-zero vorticity as the vortex core.  
Assume the fluid to be inviscid and, without any loss of generality, to have unit density.  
Taking the limit in which the cross-sectional area of the vortex core goes to zero, analogous to the planar point vortex limit, leads to a singularity in the Euler equations.  
In contrast to the planar point vortex model, the singularity cannot be avoided by simply removing the (infinite) self-induced velocity, so in practice the three-dimensional vortex ring is modeled as having a finite-sized core~\cite{Dyson1893}.   

Our discussion is focused on thin-cored circular vortex rings, as illustrated in figure \ref{fig:schematic}. 
The \emph{major radius} of the vortex ring is $R$, and its cross-section --- which we also assume to be circular to leading order --- has \emph{minor radius} $a$. 
The ratio of minor (core) radius to major radius for a vortex ring,
\begin{equation}
	\label{eq:epsilon}
	\epsilon \equiv \frac{a}{R},
\end{equation} 
which we will call the \emph{core thickness ratio}, parameterizes the relative shapes of circular vortex rings. 
In this model, all of the vorticity in the fluid is confined to the inside of this torus. 
Vortex lines are everywhere perpendicular to both the $\hat{z}$ and $\hat{r}$ axes, and ${\omega}$ is a scalar measure of the \emph{azimuthal} vorticity, i.e., $\hat{\boldsymbol{\omega}} = (0,0,\hat{\omega})$.
The circulation of the vortex ring is given by the integral $\Gamma = \int \hat{\omega}\, \mathrm{d}\hat{z}\, \mathrm{d}\hat{r}$ carried out over the cross-section of the ring in the $(\hat{z}, \hat{r})$ plane at a fixed value of $\theta$. 
Characteristic values of the quantities $R$ and $\Gamma$ can then be used to define the dimensionless system variables
\begin{equation}
	\label{eq:nondim}
	r = \frac{\hat{r}}{R}, \quad z = \frac{\hat{z}}{R}, \quad t = \frac{\hat{t}\, \Gamma}{{R}^2}, \quad \mathbf{u} = \frac{\hat{\mathbf{u}}R}{\Gamma}, \quad \omega = \frac{\hat{\omega}R^2}{\Gamma}. 
\end{equation}

It has been shown~\cite{Helmholtz1858} that an isolated circular vortex ring travels uniformly in the direction of its axis without change of shape, at a velocity $U_1$ that depends on the core thickness ratio $\epsilon$~\eqref{eq:epsilon}. 
The exact form of the velocity depends on the details of how the vorticity is distributed inside the cross-section of the ring \cite{Saffman1993}.  
A common assumption for an axisymmetric vortex ring with a circular core is to take $\omega/r = \text{constant}$, so that $\Gamma$ is independent of $a$ and $R$; we also make this assumption.  
Kelvin~\cite{Thomson1867} is attributed with determining the formula for the velocity in this case, which is, in non-dimensional form, 
\begin{equation}
	\label{eq:Kelvinformula}
	U_1(\epsilon) = \frac{1}{4 \pi} \left( \log \frac{8}{\epsilon} - \frac{1}{4} \right) + \bigo \left( \epsilon^2 \log \frac{1}{\epsilon} \right).
\end{equation}
Equation \eqref{eq:Kelvinformula} shows that as $\epsilon \to 0$, the self-induced speed of a vortex ring tends to infinity and we no longer have an admissible solution to the Euler equations. 
For finite-thickness vortex rings, the self-induced speed $U_1$ decreases with increasing $\epsilon$ (see figure \ref{U-epsilon-figure}). 
Later authors, such as Dyson~\cite{Dyson1893} and Fraenkel~\cite{Fraenkel1972}, have expanded $U_1$ in higher orders of $\epsilon$, and others have generalized \eqref{eq:Kelvinformula} to include families of vortex rings with arbitrary distributions of vorticity across the cross-section. 
There is a considerable body of work on `thick-cored' isolated vortex rings that relaxes the assumption $a \ll R$; see, for example, the family of vortex rings ranging from the infinitesimally thin-cored to the `spherical'  studied by Norbury~\cite{Norbury1973}. 

The leading-order analysis of thin-cored vortex rings, exemplified by \eqref{eq:Kelvinformula} and used throughout this work, dates back to the time of Kelvin~\cite{Thomson1867}, and has been found to give surprisingly good results when compared to theories of finite-thickness vortex rings developed by later authors. Fraenkel~\cite{Fraenkel1972} used asymptotic expansions to obtain the $\bigo (\epsilon^2 \log \epsilon^{-1})$ term in \eqref{eq:Kelvinformula}, but the latter makes up only $1\%$ of the value of $U_1$ at $\epsilon = 0.2$. 
Moreover, Norbury~\cite{Norbury1973}, in his work on truly `thick-cored' vortex rings, found that Fraenkel's asymptotic results show good agreement with the numerically computed properties of `thick-cored' vortex rings at a core thickness parameter, $\epsilon$, of $0.2$. 
Thus, we assume that our work, which is derived  from the assumption $\epsilon \ll 1$, is numerically accurate up to $\epsilon \approx 0.2$; at larger values of $\epsilon$, the assumption of a circular core is no longer valid (\cite{Saffman1993} p. 197), and the thickness ratio is properly characterized by a definition more general than \eqref{eq:epsilon}; see, e.g., equation~(2.7) in Ref.~\cite{Norbury1973}. 
For this reason, and because the higher-order terms in \eqref{eq:Kelvinformula} may no longer be negligible, our work should be considered quantitatively approximate for $\epsilon \gtrsim 0.2$.

In the frame of reference of a steadily moving, thin-cored vortex ring, one can, in general, identify three distinct regions of fluid motion: 
(1)~the core of the vortex ring, consisting of the fluid with non-zero vorticity, which moves together as one coherent body for all time; 
(2)~the potential flow surrounding the core that also maintains its identity as a coherent region moving together with the core for all time, referred to variously as the vortex `atmosphere'~\cite{Thomson1867}, the vortex `body'~\cite{Tietjens1934}, or the `vortex bubble'~\cite{Sullivan2008}; and 
(3)~the remainder of the potential flow, which extends to infinity and consists of fluid that interacts closely with the vortex ring for, at most, a finite time duration. 
This outer potential flow can be further divided into two regions: the fluid passing through the center of the vortex ring, and the fluid that passes around the outside of the vortex ring.  
As discussed in Section~\ref{singlering}, the presence of and distinction between these various regimes depends on the relative core thickness, $\epsilon$~\cite{Hicks1919}.  

Consider now two thin-cored coaxial circular vortex rings with major and minor radii $R_i$ and $a_i$, respectively, for $i=1,2$. 
When these two coaxial vortex rings interact with each other, in general their radii change in time under their mutual influence. 
By mass conservation and Helmholtz's vorticity theorems~\cite{Helmholtz1858, tait1867pm}, the quantities 
\begin{equation}\label{B-eqn} 
    B_i^2 \equiv R_i(t)\, a_i^2(t) = \text{constant}
\end{equation} 
are integrals of the motion (\cite{Dyson1893} Art. 39), remaining constant throughout any time-dependent changes in the ring and core radii $R_i$ and $a_i$. 
It is natural therefore to parameterize interacting vortex rings by the $B_i$. 
For $B_i$ to be conserved, the $a_i^2$ must vary concurrently with variations in $R_i$.  
Thus the core thickness ratio for each vortex, $\epsilon_i = a_i / R_i$, must vary as $R_i^{-3/2}$. 

In their study of two interacting vortex rings, Borisov et. al \cite{Borisov2013} examined how the \emph{relative} core thickness ratio $\beta = B_2/B_1$ changes the interaction dynamics, but the \emph{absolute} core thickness ratios $\epsilon_i$ are important parameters to consider in their own right.   
If we start with small $\epsilon_i$ and assume, as Borisov et.\ al~\cite{Borisov2013} did, that the mutual interaction of the vortex rings will be such that $a_i \ll R_i$ remains true for all time, then the vortex rings will interact as infinitely thin-cored vortex filaments. 
This assumption is qualitatively (and to some extent quantitatively) appropriate for the purpose of determining the interaction dynamics of vortex rings as long as the absolute distance between the two vortices relative to the major radii is much greater than the core thickness ratio.  
However, by not explicitly considering changes in $\epsilon_i$, this approach may not fully capture the range of possible fluid motions surrounding the vortex rings. 
The changes in $\epsilon_i$ that may occur during interaction of the vortex rings can be very large; for example, for one of the parameter values investigated by Borisov et.\ al~\cite{Borisov2013} ($\beta = 1, \gamma = 4/5, P = 2$), $\epsilon_i$ varies from $\approx 0.25$ to $\infty$ as $R_i \to 0$. 
For the flow in the vicinity of multiple interacting vortex rings, then, it is natural to conjecture that the $\epsilon_i$ will be  important parameters determining the types of possible fluid motions, even when the specific values of $\epsilon_i$  may be negligible for the purpose of modeling the vortex dynamics. 

The types of fluid motion in the vicinity of two interacting vortex rings is, in general, unsteady, and the problem does not readily admit to an analytical treatment~\cite{Bagrets1997}. 
Alternatively, a periodic array of identical axisymmetric vortex rings embedded in potential flow provides a low-order, mathematically tractable model for studying  the motion of fluid under the influence of multiple vortex rings.  
Vasilev~\cite{Vasilev1916}, and later Levy \& Forsdyke~\cite{Levy1927}, showed that each vortex in such an array moves at a constant velocity $\Uinf$ with no change in the radius of the ring, $R$. By \eqref{B-eqn}, $\epsilon$ will not change either. 
Thus, the problem can be rendered steady in an inertial frame of reference moving at the same speed as the array of vortex rings.  
It is this problem that we take as the basis for our analysis in this manuscript. 

The manuscript is organized as follows. 
In section \ref{singlering}, the known results for an isolated, thin-cored circular vortex ring are summarized, including the appearance of `thin-bodied' and `thick-bodied' vortex ring atmospheres.  
We also show, apparently for the first time, that the scaling of this system in the vicinity of the bifurcation is similar to that for a one-dimensional saddle-node bifurcation.    
In section~\ref{ringarray}, we consider a periodic array of identical, thin-cored, coaxial vortex rings. 
In section~\ref{proofs}, we give the outlines of the proof, originally due to \cite{Vasilev1916} and \cite{Levy1927}, of the existence of a stationary solution, and we present the speed of self-propagation of such an array of vortex rings.  Section~\ref{sec:bifur1} presents the `thin-body' to `thick-body' bifurcation for the periodic array, and section~\ref{sec:bifur2} discusses the bifurcation that leads to the `\connected' vortex ring system. 
Analogous to the single-ring system, these bifurcations give a scaling similar to that for a one-dimensional saddle-node bifurcation. Conclusions are presented in section~\ref{sec:conclude}.

\section{A single thin-cored, axisymmetric vortex ring} \label{singlering}

The fluid mechanics of a single axisymmetric vortex ring are well known \cite{Thomson1867,Dyson1893,Hicks1919,Saffman1993}, and we repeat the relevant analysis of that system here for context.
The equations of motion for vortex rings arise from the Euler equations of motion for an inviscid and incompressible fluid, cast in an \emph{axisymmetric} coordinate system in the vorticity-streamfunction ($\omega-\psi$) formulation. 
The axis, denoted by the $z$ direction, is aligned with the symmetry axis of the vortex ring(s), and $r>0$ extends in the radial direction, again as shown in figure~\ref{fig:schematic}. 
Generally, flows without swirl, i.e., without motion in the azimuthal direction, are considered. 
The equations of motion for the vorticity $\omega$ and the streamfunction $\psi$ are thus given in terms of the system of partial differential equations
\begin{subequations}\label{eq:euler}
\begin{align}
	\frac{\partial \omega}{\partial t} + \frac{1}{r} \frac{\partial \psi}{\partial z} \frac{\partial\omega}{\partial{r}} - \frac{1}{r} \frac{\partial \psi}{\partial r} \frac{\partial\omega}{\partial{z}} &= - \frac{1}{r^2} \frac{\partial \psi}{\partial r}\omega \frac{\partial\omega}{\partial{r}} \label{transport-eqn}, \\[1mm] 
	\frac{\partial}{\partial r} \left( \frac{1}{r} \frac{\partial \psi}{\partial r}\right) + \frac{1}{r} \frac{\partial^2 \psi}{\partial z^2} &= -\omega. \label{Poisson-eqn}
\end{align}
\end{subequations}
Given a streamfunction $\psi$ that satisfies \eqref{eq:euler}, the motion of a passive fluid particle can be determined from
\begin{equation}
	u_z = \frac{1}{r} \frac{\partial {\psi}}{\partial r}, \quad u_r = -\frac{1}{r} \frac{\partial {\psi}}{\partial z} \label{eqsmotion}.
\end{equation}

The streamfunction $\psi$ due to an infinitesimally thin vortex filament, or vortex line, with unit circulation can be derived from the Biot-Savart law, which states that the velocity induced at $\mathbf{x}$ by a vortex line $C$ parameterized by $\mathbf{y}$ is
\begin{equation}
    \mathbf{V}(\mathbf{x}) = \frac{1}{4\pi} \int_C \frac{\dy \times (\mathbf{x}-\mathbf{y})}{|(\mathbf{x}-\mathbf{y})|^3},
\end{equation}
where $\dy$ is a differential element of the vortex line $\mathbf y$, and the integral is carried out over the entire length of the vortex line $C$. 
The resulting velocity is dimensionless according to the nondimensionalization scheme in \eqref{eq:nondim}. 
For a circular vortex ring with radius one, this velocity can be expressed, in nondimensional axisymmetric cylindrical coordinates $(z,r)$, with the help of the streamfunction~\cite{Saffman1993}
\begin{equation}\label{eq:singleringpsi}
    \psi(z,r;z_0(t)) =\frac{\sqrt{r}}{2\pi} \left( \left( \frac{2}{k_0}-k_0\right) K(k_0) - \frac{2}{k_0} E(k_0) \right) \quad
    \text{with}\quad k_0 = \sqrt{ \frac{4 r}{(z - z_0(t))^2 + (r + 1)^2}}, 
\end{equation}
where $K$ and $E$ are the complete elliptic integrals of the first and second kind, respectively. 
The thickness ratio $\epsilon$ does not enter into the equation for $\psi$ explicitly because application of the Biot-Savart law is an appropriate first-order approximation for the velocity induced by a thin vortex ring at points far from the ring itself. 
This assumption was used by Hicks~\cite{Hicks1919} (p.~605) and Saffman~\cite{Saffman1993} to perform the calculations that follow.  
Higher-order expansions of \eqref{eq:singleringpsi} are available in a coordinate system centered at the vortex core center (e.g., equation~9 of section~10.2 in~\cite{Saffman1993}).
Note, however, that because the propagation speed of a vortex \emph{does} depend on $\epsilon$ to leading order, as shown in \eqref{eq:Kelvinformula}, the flow described by \eqref{eq:singleringpsi} does depend (implicitly) on $\epsilon$ through motion of the vortex center at $z_0$.

Because a vortex ring has a (time-independent) nonzero self-induced velocity $U_1$~\eqref{eq:Kelvinformula}, the axial position of the ring, $z_0$, is generally a function of time, rendering $\psi$ a time-dependent function.
A transformation to the co-moving reference frame $\zt = z - z_0 =  z - U_1 t$, with $U_1(\epsilon)$ given by \eqref{eq:Kelvinformula}, yields the steady streamfunction 
\begin{equation}
    \label{eq:singleringpsit}
    \psit (\zt,r; \epsilon) = \psi \bigl(\zt+U_1(\epsilon)\, t, r;  U_1(\epsilon)\, t\bigr)  - \frac{1}{2} U_1(\epsilon)\, r^2,
\end{equation}
for which $\zt_0 = 0$, i.e., the center of the vortex ring remains at the origin in the co-moving frame for all time. 
Isolines of $\psit$ are (steady) streamlines in a co-moving reference frame. 

\begin{figure}
\centering
\includegraphics[width=155mm]{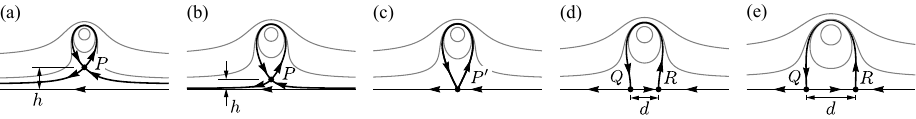}
\caption{
	Representative streamlines in the  co-moving frame of reference  for  $\epsilon =$ (a)~0.005, (b)~0.01, (c)~$\ec \approx 0.0116$, (d)~0.02, and (e)~0.05; arrows indicate direction of flow. 
	In (a,b) the vortex ring is `thin-bodied', and in (d,e) it is `thick-bodied'. 
	Axisymmetric fixed points in this co-moving frame are labeled $P, Q, R$; panel~(c) shows the special case in which there is a single co-moving point $P^{\prime}$ at $(\zt,r) = (0,0)$.
	Compare (a) and (e) with figure~10.2-2 of Ref.~\cite{Saffman1993}.
}
\label{fig:single-ring-psi}
\end{figure}

\begin{figure}
    \centering
    \includegraphics[width=155mm]{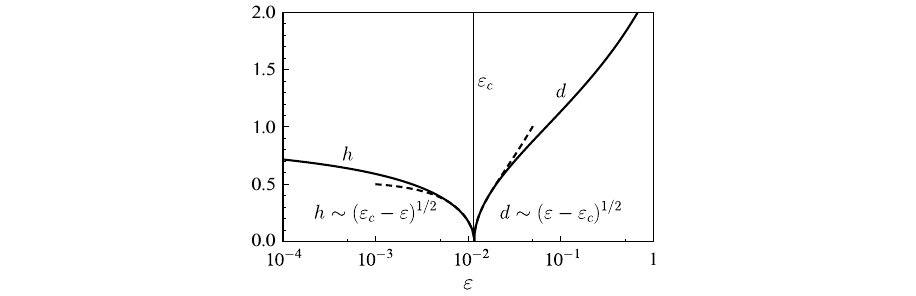}
    \caption{Variation of vertical distance $h$ and horizontal distance $d$, as labeled in figure \ref{fig:single-ring-psi}. Dashed lines show limiting behavior as $\epsilon \to \ec^-$ and $\epsilon \to \ec^+$.}
    \label{fig:singleringbifurcationdiagram}
\end{figure}

The thickness ratio $\epsilon$~\eqref{eq:epsilon} parameterizes the range of possible topologies of the co-moving fluid motion surrounding a single vortex ring~\cite{Hicks1919}. 
For sufficiently small values of $\epsilon$, there exists an axisymmetric line of fluid particles near the vortex core that moves with the same velocity as the vortex ring. 
This co-moving line appears as a stationary saddle point in the co-moving frame of reference when viewed in axisymmetric coordinates $(\zt,r)$, shown as point $P$ in figure~\ref{fig:single-ring-psi}(a,b).  
The vortex ring `atmosphere', which identifies all fluid particles that move together with the vortex core, is bounded by the axisymmetric streamline in the co-moving frame that joins the stationary saddle point to itself; that is, the atmosphere is bounded by the homoclinic orbit of the saddle point in the co-moving frame.   
The cross-section of this atmosphere in the $(\zt,r)$ plane appears as a tear-drop shape. 
Any fluid particles outside the homoclinic orbit, or separatrix, bounding the atmosphere will move away from the vortex ring as $t\to\infty$, whereas those particles inside this separatrix will remain near the vortex ring for all time. 
\begin{dfn}
\label{def:thin}
A \emph{thin-bodied} vortex ring is one whose `atmosphere' is bounded by a homoclinic orbit in the $(\zt,r)$ plane, forming a torus-shaped structure in three dimensions.
\end{dfn}
This topology has been called the \emph{toroidal configuration} by Hicks~\cite{Hicks1919}.

For large values of $\epsilon$, the co-moving fluid particles consist of two isolated points lying on center axis of symmetry, appearing as saddle points $Q, R$ in figure~\ref{fig:single-ring-psi}(d,e). 
In this case, the vortex atmosphere is bounded by the heteroclinic orbit that connects these saddle points, giving an ellipse-like region that includes a portion of the $z$ axis. 
\begin{dfn}
\label{def:thick}
A \emph{thick-bodied} vortex ring is one whose `atmosphere' is bounded by a hetereoclinic orbit in the $(\zt,r)$ plane, forming a singly-connected structure in three dimensions.
\end{dfn}
This topology has been called the \emph{singly-connected configuration} by Hicks~\cite{Hicks1919}.
It must be emphasized that it is possible for vortex rings to be `thick-bodied' under this description while continuing to be `thin-cored'; indeed, this paper is concerned exclusively with thin-cored vortex rings.

Hicks~\cite{Hicks1919} appears to have first asked the following question: is there a value of $\epsilon$ for which particles at the exact center of the vortex ring are moving with the same velocity as the ring? 
The speed $u_0$ induced by a (thin-cored) vortex ring in the axial direction along the symmetry axis is known to be~\cite{Saffman1993}
\begin{equation}
u_0 \equiv \left. \lim_{r \to 0} u_z(z,r)\right|_{z=0} = \left. \frac{1}{2} \frac{1}{(z^2 + 1)^{3/2})} \right|_{z=0} = \frac{1}{2}.
\label{u0}
\end{equation}
The point $(z,r)=(z_0,0)$ is assumed to be sufficiently far from the vortex core that the velocity induced here by the ring is independent of $\epsilon$ (to leading order). 
The self-induced velocity of the ring, $U_1(\epsilon)$~\eqref{eq:Kelvinformula}, can then be equated with \eqref{u0} to calculate a critical value of $\epsilon$, which we call $\epsilon_{c}$, for which the fluid at the center of the ring at $(z,r) = (z_0,0)$ will move with the same axial velocity as the vortex ring itself, namely
\begin{equation} 
U_1(\epsilon_c) = u_0 \implies \frac{1}{4 \pi} \left( \log \frac{8}{\epsilon_c} - \frac{1}{4} \right) = \frac{1}{2},
\label{epcrit1}
\end{equation} 
giving~\cite{Hicks1919,Saffman1993}
\begin{equation} 
\label{eq:epsilonc}
\epsilon_{c} = 8 \exp \left( -2\pi - \frac{1}{4}\right) \approx 0.0116.
\end{equation}
The co-moving streamline pattern corresponding to this critical thickness ratio is shown in figure~\ref{fig:single-ring-psi}(c).

\begin{figure}
\centering
\includegraphics[width=155mm]{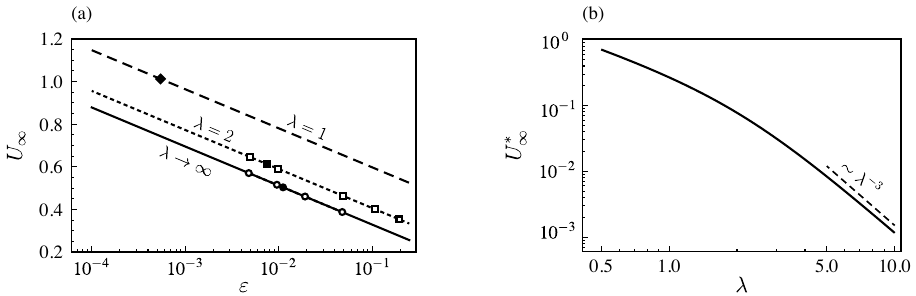}
\caption{
	(a)~Translating speed of a vortex ring array, $\Uinf(\epsilon, \lambda)$ \eqref{eq:Uinf}, for the isolated vortex ring ($\lambda \to \infty$) and two periodic arrays with $\lambda = 1,2$. 
	Circles mark the examples shown in figure~\ref{fig:single-ring-psi}; squares mark the examples shown in figure~\ref{fig:bifurcationillustration}. 
	Filled markers indicate the critical value of $\epsilon$ separating thin-bodied from thick-bodied vortex rings at the corresponding value of $\lambda$.
	(b)~Contribution to the velocity of a vortex ring induced by an infinite number of neighbors to the left and right.}
\label{U-epsilon-figure}
\end{figure}

When $\epsilon < \epsilon_{c}$, the vortex ring is `thin-bodied' and carries a torus-like atmosphere of particles around it, with a `hole' in the middle. 
When $\epsilon > \epsilon_{c}$, the vortex ring is `thick-bodied' and all of the fluid enclosed by the ring in the $(z_0, r<1, \theta)$-plane is contained in the vortex atmosphere, moving together with the ring for all time.
As $\epsilon$ increases above $\epsilon_{c}$, the thickness of the vortex atmosphere (in the $z$ direction) increases.

As shown in figure~\ref{fig:single-ring-psi}(a,b), $h$ is the distance between the central axis and the co-moving stagnation point in the plane of the ring, labeled point $P$. 
Given any value of $\epsilon < \ec$, this distance can be found by solving the equation 
\begin{equation}
    \label{eq:h-eqn}
    \left. \frac{\partial }{\partial r} \psit(\zt,r;\varepsilon)\right|_{\substack{\zt \rightarrow 0 \\ r = h}} = 0
\end{equation} 
to determine $h$. Hicks \cite{Hicks1919} was the first to give \eqref{eq:h-eqn} in terms of elliptic integrals, and he describes a procedure for solving this implicit equation (see \cite{Hicks1919}, pp.~607-611).

As shown in figure~\ref{fig:single-ring-psi}(d,e), $d$ is the distance between the heteroclinic points $Q$ and $R$ and represents the (axial) extent of the vortex atmosphere. 
The dividing streamline, given by $\psit = 0$, can be used to determine $d$ by solving 
\begin{equation}
    \label{eq:d-eqn}
    \left.\psit(d/2,r;\epsilon)\right|_{r \rightarrow 0} = 0 
\end{equation}
with any given value of $\epsilon > \ec$.  
Hicks \cite{Hicks1919}  gives a closed-form expression for the solution to \eqref{eq:d-eqn} by means of polynomial approximations to elliptic integrals; the reader is referred to \cite{Hicks1919} for this analytical treatment.

Figure \ref{fig:singleringbifurcationdiagram} shows $h$ and $d$ as functions of $\epsilon$, determined numerically from \eqref{eq:h-eqn} and \eqref{eq:d-eqn}, respectively. 
As $\epsilon$ increases from a small value up to $\ec$, $h$ decreases as $(\ec-\epsilon)^{1/2}$. 
When $\epsilon > \ec$, $d$ is zero because the homoclinic point $P$ no longer exists; instead, two heteroclinic points $Q$ and $R$ appear on the central axis, and the distance between them, $d$, varies as $(\epsilon - \ec)^{1/2}$. 
Both the bifurcation leading to the destruction of the critical point $P$ and that leading to the creation of two critical points $Q$ and $R$ therefore serve as axisymmetric analogs of the saddle-node bifurcation (see, e.g.,~\cite{Strogatz2018}). 
The destruction and creation of degenerate critical points observed here is, in fact, the expected behavior for codimension-1 bifurcations near the axis in an axisymmetric flow~\cite{Brons1999a}, which are known to scale as the square root of the bifurcation parameter.

\section{An infinite array of thin-cored, coaxial vortex rings \label{ringarray}}

Consider now an infinite array of axisymmetric, thin-cored, circular vortex rings equally spaced along the $z$ axis and having the same radius $R$. 
Each vortex ring has the same cross-sectional radius $a$, and their cross-section is assumed to be circular to leading order in $\epsilon=a/R$. 
Let the distance between two successive vortex rings along the $\hat z$ axis be $L$, and define $\lambda \equiv L/R$. 
Suppose also that $L \gg a$ (i.e., $\lambda \gg \epsilon$), so that the mutual influence of the vortices can be modeled accurately by infinitesimally thin vortex filaments interacting with each other. 
The vortices are embedded in an inviscid fluid that fills all of space and obeys Euler's equations in axisymmetric form \eqref{eq:euler}.

\subsection{Existence of a stationary solution \label{proofs}}

\begin{prp}
Equations~\eqref{eq:euler} admit the following stationary solution
\begin{equation}
    \label{relative-psi}
    \psiinft(\zt,r; \epsilon, \lambda) = \psiinf(z,r; \lambda) - \frac{1}{2}\, \Uinf(\epsilon, \lambda)\, r^2,
\end{equation}
with $z=\zt +\Uinf t$, where $\psiinf$ is the streamfunction due to an infinite array of equally spaced, identical, coaxial, infinitesimally thin circular vortex filaments of unit intensity, and $\Uinf(\epsilon, \lambda)$ is the (constant) speed of translation, in the axial direction, of the infinite system of vortex rings. 
The function $\psiinf$ can be constructed from a Green's function as
\begin{subequations}\label{absolute-psi}
\begin{gather}
    \label{psiinf-def}
    \psiinf(z,r; \lambda) = \int_{\Omega} \Ginf(z,r;\zb,\rb,\lambda)\,\omega(\zb,\rb)\, d\zb\,d\rb, \\
    \label{G-def}
    \Ginf(z,r;\zb,\rb,\lambda) = \sum^{+\infty}_{j=-\infty} G(z,r;\zb+j\lambda,\rb),
\end{gather}
where 
\begin{equation}
	G(z,r;\zeta,\rb) = \frac{\sqrt{r \,\rb}}{2\pi} \left( \left( \frac{2}{k}-k\right) K(k) - \frac{2}{k}\, E(k) \right) \quad \text{with}\quad k = \left( \frac{4 r \,\rb}{(z - \zeta)^2 + (r + \rb)^2}\right)^{1/2};
\end{equation}
\end{subequations}
the functions $K$ and $E$ are the complete elliptic integrals of the first and second kind, respectively. 
The function $\psiinf$ is defined on the periodic strip $\Omega = \{(z,r) : (z- z_0(t)) \in (-\lambda/2, +\lambda/2) , r > 0 \}$. 
For circular, infinitely thin-cored axisymmetric rings, the vorticity distribution $\omega(z, r)$ can be approximated as the two-dimensional delta function $\delta(z-z_0,r-1)$, in which case
\begin{equation}
	\psiinf (z,r; \lambda) = \Ginf (z,r; z_0, 1, \lambda).
\end{equation}
\end{prp}

\begin{proof}
A full proof was given by Vasilev \cite{Vasilev1916} and Levy \& Forsdyke \cite{Levy1927}. 
Here, we sketch the outlines of the proof for reference; the reader is referred to the original works \cite{Vasilev1916,Levy1927} for a complete discussion.
\end{proof}

\begin{lem}
The infinite series in Eq.~\eqref{G-def} is convergent in the periodic strip $\Omega$ except at the singular point $(z_0,1)$ for the case $\omega(z,r)=\delta(z-z_0,r-1)$.
\end{lem}

This Lemma follows from the fact that the streamfunction due to a single infinitesimally thin-cored vortex ring is well-defined everywhere except at the ring itself, and that the additional contribution from each successive vortex ring in the summation (which are all outside $\Omega$) decays as~$1/z$.

\begin{lem}
A system consisting of a periodic array of identical, coaxial, axisymmetric, infinitesimally thin circular vortex rings translates uniformly with a constant, finite, speed $\Uinf(\epsilon, \lambda)$ in the axial direction.
\end{lem}

This Lemma can be shown using direct calculation. 
Let
\begin{equation}\label{eq:Uinf}
	\Uinf(\epsilon, \lambda) = U_1(\epsilon) + \Uinfd(\lambda),
\end{equation}
where $U_1(\epsilon)$ is the translating speed of a single vortex given by \eqref{eq:Kelvinformula} and $\Uinfd(\lambda)$ is the contribution to the translational speed from the remaining vortex rings outside $\Omega$. 
To calculate $\Uinfd$, consider a vortex ring with unit radius located at the axial position $z_0$. 
Then, the instantaneous streamfunction governing the motion of this vortex ring will be given by the \emph{deleted} Green's function $\Gd$, in which the $j=0$ term is dropped from \eqref{G-def}, i.e.,
\begin{equation}
	\label{eq:Gdeleted}
	\psid (z,r; \lambda) = \Gd(z,r;z_0,1,\lambda).
\end{equation}
The axial velocity induced at $(z_0,1)$ is then given by the radial derivative of the deleted Green's function or, equivalently, by a direct summation of infinite Biot-Savart integrals evaluated at one of the vortex rings' locations \cite{Vasilev1916, Levy1927}, resulting in
\begin{align}
	\Uinfd(\lambda)= \left.\frac{1}{r}\frac{\partial \psid}{\partial r} \right|_{r=1,z=z_0} &= \left. \frac{\partial }{\partial r} \Gd(z_0, r; z_0,1,\lambda) \right|_{r = 1} \notag \\
	&= \frac{1}{2 \pi}\sum^{\infty}_{n=1} \int^{2\pi}_{0} \frac{1 - \cos \phi}{\left(2 (1 - \cos \phi) + n^2 \lambda^2 \right)^{3/2}} \ d \phi.  \label{eq:Uinfd}
\end{align}
The variation of $\Uinfd(\lambda)$ with $\lambda$ is shown in figure~\ref{U-epsilon-figure}(b), and the resulting influence on $\Uinf(\epsilon,\lambda)$ is shown in figure~\ref{U-epsilon-figure}(a).  
For $\lambda \gg 1$, the contribution to a vortex ring's velocity from its neighbors is small, and we have the limit
\begin{equation}
	\lim_{\lambda \to \infty} \Uinfd(\lambda) = \zeta(3)\,\lambda^{-3},
\end{equation}
where $\zeta(n)$ is the Riemann zeta function.  

The expression for $\Uinfd(\lambda)$ in \eqref{eq:Uinfd} was found independently by Vasilev~\cite{Vasilev1916} and Levy~\& Forsdyke~\cite{Levy1927}. 
In practice, the infinite sum can be approximated by a truncated series in which the influence of pairs of vortex rings is symmetrically added, i.e., the deleted Green's function appearing in \eqref{eq:Gdeleted} can be approximated as 
\begin{equation}
	G^{\ast}_N(z,r;z_0,1,\lambda) = \sideset{}{'} \sum^{+N}_{j=-N} G(z,r;z_0+j\lambda,1);
\end{equation}
the prime on the summation indicates omission of the term $j=0$.
The approximated velocity $\Uinfd(\lambda; N)$ converges quickly with the number of pairs $N$ included in the truncated sum for all but the smallest values of $\lambda$. 
In this work, a convergence study was conducted and it was found that increasing $N > 500$ did not change any of the results meaningfully. 
The convergence is very rapid for $\lambda \gtrsim 0.25$; for lower values of $\lambda$, more sophisticated methods of computing periodic Green's functions, such as those reported in \cite{Oroskar2006}, could be used instead.

\begin{lem}
A periodic array of identical, coaxial, axisymmetric, infinitesimally thin circular vortex rings propagate with no change in their radii.
\end{lem}

The radial velocity of one of the vortex rings is determined by evaluating the limit of \eqref{eqsmotion},
\begin{equation}
	u_r(z_0,1) = -\frac{1}{r} \left. \frac{\partial \psid}{\partial z} \right|_{z \rightarrow z_0, r \rightarrow 1} = -\frac{1}{r} \left. \frac{\partial}{\partial z} \right|_{z \rightarrow z_0} \hspace*{-7mm}\Gd (z,r; z_0,1,\lambda).
\end{equation}
Only the deleted Green's function \eqref{eq:Gdeleted} is involved because the self-induced radial velocity from the ring centered at $z=z_0$ is zero, corresponding in \eqref{G-def} to the term with $j=0$.  
The contributions from the remaining terms cancel pairwise, 
\begin{equation}
	\left. \frac{\partial}{\partial z} \right|_{z \rightarrow z_0} \hspace*{-7mm} G(z,r; z_0+j\lambda,1) + \left. \frac{\partial}{\partial z} \right|_{z \rightarrow z_0} \hspace*{-7mm} G(z,r; z_0-j\lambda,1) = 0 \qquad \forall\, i \in \mathbb{N},
\end{equation}
and there is no radial component of velocity at the location of the vortex ring core.   
Because each vortex ring is indistinguishable from the others, this holds true for all vortices in the infinite array. 
Hence, there is no change in the radii of the vortex rings. 

\begin{figure}
\centering
\includegraphics[width=155mm]{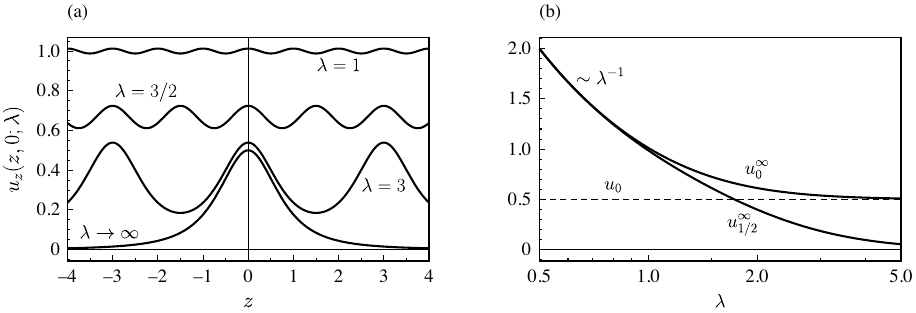}
\caption{
	Axial speed (in the absolute frame of reference) induced at the central axis $r=0$. 
	(a)~Speed as a function of position along the $z$ axis for a vortex ring array centered instantaneously at $z=0$ for $\lambda = 1, 2.5, 3$ and $\lambda \to \infty$. 
	(b)~Speed induced by a periodic array of vortex rings at the center of a vortex ring, $u_0^{\infty}$ \eqref{eq:u0inf}, and at the midpoint (along the axis) between two successive rings, $u_{1/2}^{\infty}$ \eqref{eq:u12inf}, as a function of $\lambda$. 
	The dashed line shows $u_0$, the axial centerline speed induced by a single vortex ring.}
	\label{fig:centerlinevel}
\end{figure}

The axial velocity induced by an array of vortex rings on their common axis $z$ is given instantaneously by 
\begin{equation}
	\label{eq:uzinf}
	u_z^{}(z,0;\lambda) = \lim_{r \to 0} \frac{\partial \psiinf}{\partial r}= \sum^{n=+\infty}_{n=-\infty} \frac{1}{2(1+z^2 - 2 n \lambda z + n^2 \lambda^2)^{3/2}};
\end{equation}
the expression on the right-hand-side can be derived by considering an appropriate limit of the Biot-Savart law \cite{Saffman1993} for an infinitesimal filament. 
The dependence of this axial centerline velocity on the array spacing is shown in figure~\ref{fig:centerlinevel}(a).  
As $\lambda$ decreases from the $\lambda\rightarrow\infty$ isolated ring limit, the mean of $u_z^{}(z,0;\lambda)$ increases while the deviation from that mean decreases and becomes more sinusoidal.

\subsection{`Thin-bodied' and `thick-bodied' arrays}
\label{sec:bifur1}

\begin{figure}
\centering
\includegraphics[width=155mm]{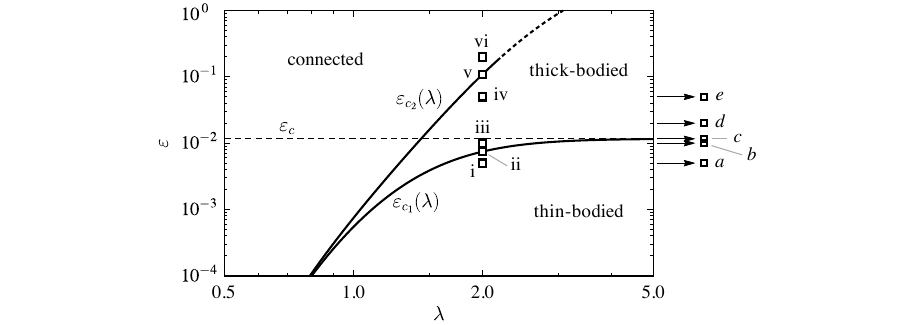}
\caption{
	Bifurcations in the streamline topology in the  $(\lambda,\epsilon)$ parameter space.  
	The appearance of `thin-bodied', `thick-bodied', and `\connected' atmospheres are delineated by the curves $\econe(\lambda)$~\eqref{eq:epsilonc1} and $\ectwo(\lambda)$~\eqref{epsilonc2}; $\ectwo$ is dashed for $\epsilon \gtrsim 0.2$ to indicate that the `thin-cored' assumption becomes inaccurate for large $\epsilon$. 
	Points i--vi correspond to the streamline patterns shown in figure~\ref{fig:bifurcationillustration} for $\lambda=2$; 
points labeled $a$--$e$ correspond to the streamline patterns shown in figure~\ref{fig:single-ring-psi} for which $\lambda\to\infty$. 
	The horizontal dashed  line shows $\ec$~\eqref{eq:epsilonc}.
}
\label{fig:parameterspace}
\end{figure}

As discussed in section~\ref{singlering}, the atmosphere of a single thin-cored axisymmetric vortex ring can have two distinct topologies in the co-moving frame --- `thin-bodied' or `thick-bodied' --- as a function of the core thickness ratio $\epsilon$, with the bifurcation in the topology occurring for $\epsilon = \epsilon_c$.  
This topological bifurcation also occurs for a periodic array of vortex rings, but in this system the critical value of the core thickness ratio is a function of the spacing ratio, $\lambda$.  
This departure from the isolated ring behavior comes as a result of two changes.  
First, the propagation speed of a periodic vortex ring array is given not just by $U_1$~\eqref{eq:Kelvinformula} but by $\Uinf(\epsilon,\lambda) = U_1(\epsilon) + \Uinfd(\lambda)$, with $\Uinfd$ defined in \eqref{eq:Uinfd}. 
Second, the axial speed at the center of one of the vortex rings, $(\zt,r)=(j \lambda,0) \, \forall j \in \ints$, is no longer given by~\eqref{u0}, because the influence of the ring's neighbors needs to be taken into account. 
The axial speed induced by an array of vortex rings at $(\zt,r) = (j \lambda ,0) \, \forall j \in \ints$ can be found by substituting $z=0$ in \eqref{eq:uzinf}, which gives
\begin{equation}
	\label{eq:u0inf}
	u_0^{\infty}(\lambda) \equiv u_z(0,0;\lambda) = \sum _{n=-\infty }^{\infty } \frac{1}{2 \left(1+n^2 \lambda^2\right)^{3/2}}.
\end{equation}
The values of the parameters $(\lambda,\epsilon)$ for which a fluid particle at the center of one of the vortex rings is stationary (in a frame of reference moving with speed~$\Uinf$) is given by solutions to the equation ${\Uinf}\left( \lambda, \econe \right) = u_0^{\infty} \left( \lambda \right)$ which, by combining \eqref{eq:Kelvinformula}, \eqref{eq:Uinfd}, and \eqref{eq:u0inf}, leads to
\begin{equation}
	\econe(\lambda) = 8 \exp \left(- 4 \pi \left( u_0^{\infty}(\lambda) - \Uinfd(\lambda) \right) - \frac{1}{4} \right).
	\label{eq:epsilonc1}
\end{equation}
As shown in figure~\ref{fig:parameterspace}, this $\lambda$-dependent curve delineates the appearance of `thin-bodied' and `thick-bodied' streamline topologies in the period vortex ring array. 
At large values of $\lambda$, the effect of neighboring vortex rings attenuates because $\Uinfd \to 0$, and as a result $\econe \to \ec$ as $\lambda \to \infty$.

\begin{figure}
\centering
\includegraphics[width=155mm]{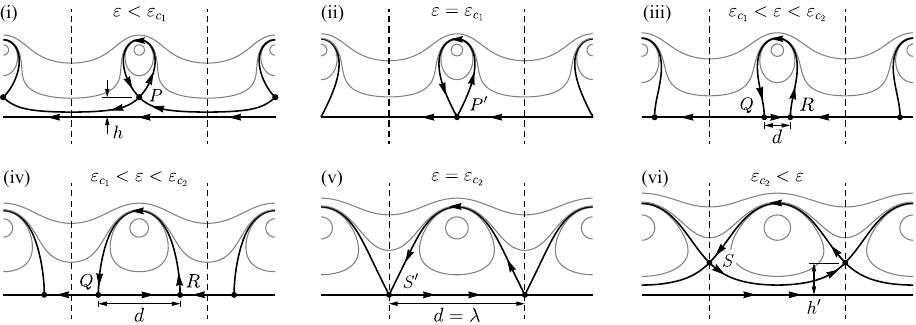}
\caption{Representative streamlines in the co-moving frame of reference for $\lambda=2$ and $\epsilon = $ (i)~0.005, (ii)~$\econe(2)\approx 0.00745$, (iii)~0.01, (iv)~0.05, (v)~$\ectwo(2)\approx 0.109$, and (vi)~0.2;  arrows indicate direction of flow. 
In (i) the vortex ring is `thin-bodied', in (iii, iv) it is `thick-bodied', and in (vi) is it `\connected'. Panels~(ii, v) show the special cases in which there is a single co-moving point at  (ii)~$(\zt,r) = (0,0)$ and (v)~$(\zt,r) = (\lambda/2,0)$, respectively.
}
\label{fig:bifurcationillustration}
\end{figure}

Figure~\ref{fig:bifurcationillustration}(i)--(iii) illustrate the changes in the topology of streamlines in the co-moving frame of reference as $\epsilon$ is increased through $\econe$.
For $\epsilon < \econe$, the atmosphere of each vortex ring is `thin-bodied' according to Definition~\ref{def:thin}.  
The homoclinic point $P$ is located a vertical distance $h$ from the centerline axis, and the fluid near this axis remains relatively undisturbed, `left behind' by the passage of vortex rings. 
At the critical value $\epsilon = \econe(\lambda)$, the homoclinic orbit bounding the vortex atmosphere touches the $z$ axis at a single point, $P^{\prime}$.  
As $\epsilon$ increases from $\econe(\lambda)$, each vortex ring becomes `thick-bodied' according to Definition~\ref{def:thick}, with a vortex atmosphere bounded by a simply-connected, ellipse-like surface that encompasses a (finite) portion of the $z$ axis.   

\begin{figure}
\centering
\includegraphics[width=155mm]{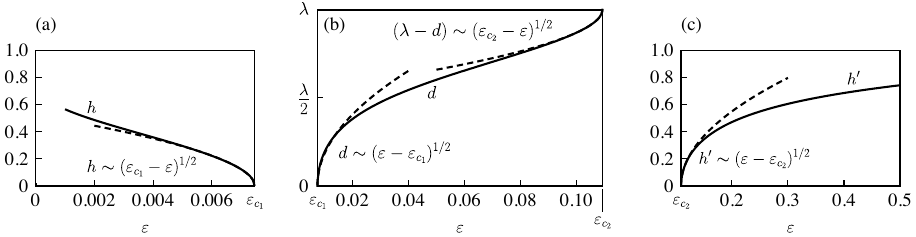}
\caption{
	Horizontal and vertical spacings of the co-moving points $P$, $Q$, $R$, and $S$ for the case $\lambda = 2$ shown in figure~\ref{fig:bifurcationillustration}. 
	Dashed lines give the scaling laws indicated in each panel; solid lines show numerically computed values for (a)~$h$, the vertical position of $P$, (b)~$d$, the horizontal distance between $Q$ and $R$, and (c)~$h'$, the vertical position of $S$.  Note that the horizontal axis runs continuously across all three panels from $\epsilon = 0$ to~$0.5$.
}
\label{fig:bifurcationdiagram}
\end{figure}

In the neighborhood of the critical point, the location(s) of the saddle point(s) again exhibit scalings like that for a saddle-node bifurcation, as noted in section~\ref{singlering} for the isolated vortex ring.  
As $\epsilon \to \econe^-(\lambda)$, the point~$P$ approaches the centerline axis asymptotically as $h \sim (\econe - \epsilon)^{1/2}$, as shown in figure~\ref{fig:bifurcationdiagram}.  
If instead $\epsilon \to \econe^+(\lambda)$, the two saddle points at $Q$ and $R$ approach each other asymptotically as $d \sim (\epsilon - \econe)^{1/2}$.  
Locally, $Q$ looks like a one-dimensional `saddle' and $R$ a one-dimensional `node' on the $z$ axis.  

Qualitatively, this  change in co-moving streamline topology with varying $\epsilon$ is similar to that shown in figure~\ref{fig:single-ring-psi} for the isolated vortex ring, but the details of the streamline pattern now depend on the periodic spacing~$\lambda$.  
In the isolated-vortex limit, $\lambda \to \infty$, we find that $\econe \to \ec$, which is to be expected because, at large $\lambda$, the effect of $\Uinfd$ attenuates to zero and $u_0^{\infty}$ tends to $1/2$, which means that \eqref{eq:epsilonc1} tends to \eqref{eq:epsilonc}. 
In the opposite limit, $\lambda \to 0$, the value of $\econe$ drops off steeply, as shown in fig.~\ref{fig:parameterspace},  revealing that for closely-spaced vortex ring arrays the threshold for a ring to be `thick-bodied' is low and can be easily attained by an array of vortex rings even with $\epsilon\ll 1$.  
For example, the thick-bodied vortex ring array in figure~\ref{fig:bifurcationillustration}(iii) and the isolated thin-bodied  vortex ring in figure~\ref{fig:single-ring-psi}(b) both have $\epsilon = 0.01$.

\subsection{`Connected' arrays}
\label{sec:bifur2}

For a periodic array of vortex rings, a new topology arises that cannot occur for an isolated vortex. 
As $\epsilon$ increases from $\epsilon>\econe$, such as shown in figure~\ref{fig:bifurcationillustration}(iii, iv), the width of the atmosphere of each ring increases in the $z$-direction.   
At a new critical value of the thickness ratio, $\epsilon = \ectwo(\lambda)$, the atmospheres of adjacent vortex rings touch at their midpoint along the $z$ axis, as shown in figure~\ref{fig:bifurcationillustration}(v).  
In this special case, \emph{every} fluid particle along the $z$ axis is contained within in the atmosphere of a vortex ring, and the point(s) labeled $S'$ at $(\zt,r)=(\lambda/2+j\lambda,0)\, \forall j \in \mathbb{Z}$ is the only stationary point in the co-moving frame. 
As $\epsilon$ further increases for $\epsilon > \ectwo(\lambda)$, the stationary point in the $(\zt,r)$ plane lifts from the central axis of symmetry to the point $S$ at a position $r=h'$, as illustrated in figure~\ref{fig:bifurcationillustration}(vi). 
Similar to the `thin-bodied' case, this stationary point corresponds to an axisymmetric co-moving line of fluid particles. 
In contrast to the thin-bodied case, the atmospheres of neighboring vortex rings touch, or are `\connected', at this point.  
The vortex ring atmosphere is now bounded by heteroclinic orbits, that is, by the streamlines, or separatrices, that join the stationary saddle point at $S$ with a neighboring periodic image of that point.  

\begin{dfn}
\label{def:joined}
A \emph{\connected} array of vortex rings is one in which the `atmospheres' of successive vortex rings are bounded by hetereoclinic orbits that are joined mid-way between the vortex rings at a saddle point.
\end{dfn}

This second critical value, $\ectwo > \econe$, occurs when the speed of the vortex rings, $U_{\infty}$, is matched by the axial velocity induced at the point $(z,r) = (z_0+\lambda/2,0)$ (and its periodic images), which can be found by substituting $z = \lambda/2$ in  \eqref{eq:uzinf}, giving
\begin{equation}
	\label{eq:u12inf}
    u_{1/2}^{\infty}(\lambda) \equiv u_z(1/2,0;\lambda) 
    = \sum _{n=-\infty }^{\infty } \frac{4}{\left(4+(1-2 n)^2 \lambda^2\right)^{3/2}} .
\end{equation}
Solving ${\Uinf} \left( \lambda, \ectwo \right) = u_{1/2}^{\infty} \left( \lambda \right)$ by combining \eqref{eq:Kelvinformula}, \eqref{eq:Uinfd},  and \eqref{eq:u12inf} gives
\begin{equation}
	\ectwo(\lambda) = 8 \exp \left(-4 \pi \left( u_{1/2}^{\infty}(\lambda) - \Uinfd(\lambda) \right) - \frac{1}{4} \right),
	\label{epsilonc2}
\end{equation}
which is shown in fig.~\ref{fig:parameterspace}. 

The bifurcation at $\epsilon = \ectwo(\lambda)$ also exhibits a scaling similar to that for a saddle-node bifurcation.  
As $\epsilon \to \ectwo^-(\lambda)$, the points~$Q$ and $R$ approach each other along the $z$ axis with a horizontal spacing that decreases asymptotically as $d \sim (\ectwo - \epsilon)^{1/2}$, as shown in figure~\ref{fig:bifurcationdiagram}. 
As $\epsilon \to \econe^+(\lambda)$, the point $S$ approaches the centerline axis asymptotically as $h' \sim (\epsilon - \ectwo)^{1/2}$.    

As with the transition from thin- to thick-bodied vortex ring arrays, the appearance of \connected\ vortex ring arrays depends on the periodic spacing~$\lambda$. 
In the isolated-vortex limit $\lambda \to \infty$, we find that $\ectwo \to 8/e^{1/4}$, far above the `small' values needed for our the `thin-cored' assumption to hold; practically, this means that for large $\lambda$, the bifurcation $\ectwo$ cannot be reached by thin-cored vortex rings. 
This behavior is to be expected because the `midpoint' of two vortex rings is infinitely far away when $\lambda \to \infty$, and an isolated ring cannot carry all fluid particles along its axis because it would need an infinitely large atmosphere to do so. 
In the opposite limit, $\lambda \to 0$, $\ectwo$ drops off steeply, scaling at low $\lambda$ in a manner identical to that of $\econe$. 
Mathematically it can be seen that in the limit of very closely spaced vortex ring arrays, the right hand sides of \eqref{eq:epsilonc1} and \eqref{epsilonc2} approach each other and become nearly identical. 
Physically, it should be expected that for closely-spaced vortex rings, the threshold value of $\epsilon$ at which the atmospheres of neighboring vortex rings will touch is very low.  
As a consequence, for $\lambda<1$ vortex ring arrays make a very rapid transition from `thin-bodied' to `\connected' with increasing $\epsilon$, with `thick-bodied' rings occurring for a vanishingly small range $\epsilon \in (\econe, \ectwo)$ for $\lambda\lesssim 0.8$.  

For thin-cored vortex ring arrays, the speed of the fluid along the central axis of symmetry, $u_z(z,0;\lambda)$ \eqref{eq:uzinf}, becomes larger and increasingly uniform as $\lambda\to 0$, as illustrated in fig.~\ref{fig:centerlinevel}. 
Under the thin-cored assumption, this speed is, to leading order, independent of the relative core thickness $\epsilon\ll1$. 
On the other hand, the translating speed of the vortex ring array, $\Uinf(\epsilon,\lambda)$ \eqref{eq:Uinf}, decreases with increasing $\epsilon$, as shown in fig.~\ref{U-epsilon-figure}. 
For `\connected' vortex ring arrays, the value of $\epsilon$ is large enough or, correspondingly, the value of $\lambda$ is small enough that there exists a `tube' of particles around the $z$ axis that are swept \emph{forward} in the co-moving frame with an average speed that is greater than that of the vortex ring.  

It should be emphasized that the bifurcations being considered here are bifurcations in the streamline topology, i.e.~they are bifurcations of solutions to the ordinary differential equations~\eqref{eqsmotion}, not of the partial differential equations~\eqref{eq:euler}. 
Indeed, the two bifurcations that occur for axi-symmetric, thin-cored vortex ring arrays are the simplest ones that can occur near the axis in an axisymmetric flow~\cite{Brons2021}. 
Specifically, these two bifurcations correspond to the creation and destruction of degenerate critical points in an axisymmetric flow for the 2nd order normal form when \eqref{eqsmotion} are expanded in a Taylor series about the origin~\cite{Brons1999a}. 
Because this set of bifurcations is codimension-1 (i.e., they can be achieved by variation of a single parameter $\epsilon$ while keeping $\lambda$ fixed), we expect to see features from the normal form of order $N=2$ and no higher. 
The same bifurcations in streamline topology have been observed in numerical simulations of the flow inside a cylinder with counter-rotating ends~\cite{Brons1999}.

\section{Conclusions} \label{sec:conclude}

We have considered the motion due to a periodic array of identical, coaxial, axisymmetric, infinitesimally thin circular vortex rings in an unbounded inviscid fluid as a function of two non-dimensional parameters: $\lambda$, the ratio of their axial separation to their radii, and $\epsilon$, their common `thickness ratio'. 
This system is, in a sense, the simplest problem involving the `interaction' of multiple vortex rings, where the radii, separation, and thickness of the rings --- the variables that generally characterize an $N$-vortex ring problem --- all remain constant for all time.
The interaction dynamics of the vortex rings themselves are therefore quite simple, and have been known for more than a century~\cite{Vasilev1916,Levy1927}. 
However, the motion of the surrounding fluid in such a system, or indeed for any system of interacting coaxial vortex rings, has received significantly less attention. 
The fact that this problem can be viewed as steady in a uniformly-translating reference frame renders it mathematically tractable and makes it useful for investigating how the separation and thickness of vortex rings affect fluid transport in their vicinity.

As is known for a single vortex ring, each vortex in the periodic array carries with itself an `atmosphere' of particles whose size and topology is modulated by $\epsilon$.  
For $\epsilon < \econe$, the vortex rings are `thin-bodied'.  
The bounding value $\econe$ is a strong function of the separation distance between neighboring vortex rings, $\lambda$.  
For $\lambda\gg 1$, $\econe(\lambda)\approx\ec$ as is expected, but $\econe(\lambda)$ decreases rapidly with decreasing $\lambda$, with $\econe(\lambda)\ll\ec$ for $\lambda=\bigo(1)$.  
For $\econe<\epsilon<\ectwo$, the vortex rings are `thick-bodied', with the range of $\epsilon$ over which these `thick-bodied' rings are found decreasing rapidly with decreasing $\lambda$.  
The new, higher critical value, $\ectwo(\lambda)$, is the thickness ratio value at which the mid-point along the central axis  between neighboring vortex rings becomes a fixed point in the co-moving frame of reference. 
These critical values, $\econe$ and $\ectwo$, occur on two codimension-1 bifurcation curves that meet asymptotically at $(\lambda,\epsilon) = (0,0)$ and are the simplest possible bifurcations that can occur in an axisymmetric system near the axis, as first observed in Ref.~\cite{Brons1999}.  
For $\epsilon>\ectwo(\lambda)$, the streamline topology in the co-moving frame gives what we refer to as `\connected' vortex rings, for which the atmospheres of two neighboring vortex rings meet at a circle of co-moving fluid particles that is located mid-way between the two rings.  
For `thin-bodied' rings, the rings travel faster than the fluid near the central axis of symmetry; for `thick-bodied' rings, fluid along a portion of the central axis is contained within the atmosphere of each ring and thus moves with the rings; and for `\connected' rings the fluid near the central axis travels \emph{faster} than the vortex rings.  
An interesting feature of this system is that, for a fixed value of thickness ratio $\epsilon < \ec$, a periodic array of thin-cored, axisymmetric vortex rings may be thin-bodied, thick-bodied, or \connected, depending on the value of the inter-vortex spacing,~$\lambda$.

In practice, it is possible to find an approximation of a periodic array of axisymmetric vortex rings using pulsed jets \cite{Krueger2005,Krueger2009} or observing the wakes of marine animals~\cite{Gordon2017}. 
From the perspective of vortex roll-up for a single ring~\cite{Gharib1998}, Linden \&~Turner have proposed an `optimum formation length' for the formation of vortex rings from a pulsed jet~\cite{Linden2001}.  
Our results suggest that, depending upon the desired impact of the vortex rings on the surrounding fluid, one might also need to consider the effect of the vortex spacing and core size on the topology of the vortex ring atmosphere; that is, one might consider if the `optimum formation length' leads to vortex rings that are `thin-bodied', `thick-bodied', or `\connected'.

\section*{Acknowledgements}

The authors wish to thank Dr.~Shane Ross and Dr.~Morten Br{\o}ns for their helpful discussions regarding bifurcations in the topology of streamlines, as well as the anonymous reviewers for their constructive feedback, particularly for pointing out the close connection between our work and that of Hicks~\cite{Hicks1919}.

\section*{Funding}

This research was funded in part by the National Science Foundation's Graduate Research Fellowship under Grant No.~1840995.

\section*{Conflict of interest}

The authors declare that they have no conflicts of interest.

\section*{Authors' contributions}

E.M. conceived of the core ideas of the manuscript with guidance from M.A.S. Mathematical derivations and numerical calculations were carried out by E.M. Both authors contributed equally to the writing of the text. 

\bibliographystyle{abbrv}

\begin{thebibliography}{10}

\bibitem{Allen2007}
J.~J. Allen, Y.~Jouanne, and B.~N. Shashikanth.
\newblock {Vortex interaction with a moving sphere}.
\newblock {\em Journal of Fluid Mechanics}, 587:337--346, sep 2007.


\bibitem{Bagrets1997}
A.~A. Bagrets and D.~A. Bagrets.
\newblock {Nonintegrability of two problems in vortex dynamics}.
\newblock {\em Chaos: An Interdisciplinary Journal of Nonlinear Science},
  7(3):368--375, sep 1997.

\bibitem{Bearman1988}
P.~W. Bearman and M.~Takamoto.
\newblock {Vortex shedding behind rings and discs}.
\newblock {\em Fluid Dynamics Research}, 3:214--218, 1988.


\bibitem{Borisov2013}
A.~V. Borisov, A.~A. Kilin, and I.~S. Mamaev.
\newblock {The dynamics of vortex rings: Leapfrogging, choreographies and the
  stability problem}.
\newblock {\em Regular and Chaotic Dynamics}, 18(1-2):33--62, jan 2013.

\bibitem{Brons1999a}
M.~Br{\o}ns.
\newblock {Streamline Topology of Axisymmetric Flows}.
\newblock In A.~N. {S{\o}rensen J.N., Hopfinger E.J.}, editor, {\em Simulation
  and Identification of Organized Structures in Flows}, pages 213--222.
  Springer, 1999.

\bibitem{Brons1999}
M.~Br{\o}ns, L.~K. Voigt, and J.~N. S{\o}rensen.
\newblock {Streamline topology of steady axisymmetric vortex breakdown in a
  cylinder with co- and counter-rotating end-covers}.
\newblock {\em Journal of Fluid Mechanics}, 401:275--292, dec 1999.

\bibitem{Brons2021}
M.~Br{\o}ns.
\newblock {Private communication, November 2021}.
\newblock In A.~N. {S{\o}rensen J.N., Hopfinger E.J.}, editor, {\em Simulation
  and Identification of Organized Structures in Flows}, pages 213--222.
  Springer, 1999.

\bibitem{Dabiri2005}
J.~O. Dabiri, S.~P. Colin, J.~H. Costello, and M.~Gharib.
\newblock {Flow patterns generated by oblate medusan jellyfish: field
  measurements and laboratory analyses}.
\newblock {\em Journal of Experimental Biology}, 208(7):1257--1265, apr 2005.

\bibitem{Dyson1893}
F.~W. Dyson.
\newblock {The Potential of an Anchor Ring. Part II}.
\newblock {\em Philosophical Transactions of the Royal Society A: Mathematical,
  Physical and Engineering Sciences}, 184:1041--1106, 1893.

\bibitem{Ellington1984}
C.~P. Ellington.
\newblock {The aerodynamics of hovering insect flight. V. A vortex theory}.
\newblock {\em Philosophical Transactions of the Royal Society of London. B,
  Biological Sciences}, 305(1122):115--144, feb 1984.

\bibitem{Fraenkel1972}
L.~E. Fraenkel.
\newblock {Examples of steady vortex rings of small cross-section in an ideal
  fluid}.
\newblock {\em Journal of Fluid Mechanics}, 51(1):119--135, jan 1972.

\bibitem{Gharib1998}
M.~Gharib, E.~Rambod, and K.~Shariff.
\newblock {A universal time scale for vortex ring formation}.
\newblock {\em Journal of Fluid Mechanics}, 360:121--140, apr 1998.

\bibitem{Gordon2017}
M.~S. Gordon, R.~Blickhan, J.~O. Dabiri, and J.~J. Videler.
\newblock {\em {Animal locomotion: Physical principles and adaptations}}.
\newblock Taylor $\backslash${\&} Francis Group, 2017.

\bibitem{Heinzel1987}
H.-G. Heinzel and M.~Dambach.
\newblock {Travelling air vortex rings as potential communication signals in a
  cricket}.
\newblock {\em Journal of Comparative Physiology A 1987 160:1}, 160(1):79--88,
  jan 1987.

\bibitem{Helmholtz1858}
H.~Helmholtz.
\newblock {{\"{U}}ber Integrale der hydrodynamischen Gleichungen, welche den
  Wirbelbewegungen entsprechen}.
\newblock {\em Journal fur die Reine und Angewandte Mathematik},
  1858(55):25--55, 1858.

\bibitem{Hicks1919}
W.~Hicks.
\newblock { LIX. The mass carried forward by a vortex }.
\newblock {\em The London, Edinburgh, and Dublin Philosophical Magazine and
  Journal of Science}, 38(227):597--612, 1919.

\bibitem{Hicks1922}
W.~M. Hicks.
\newblock {On the mutual threading of vortex rings}.
\newblock {\em Proceedings of the Royal Society of London. Series A, Containing
  Papers of a Mathematical and Physical Character}, 102(715):111--131, nov
  1922.


\bibitem{Krueger2009}
P.~S. Krueger.
\newblock {Vortex ring velocity and minimum separation in an infinite train of
  vortex rings generated by a fully pulsed jet}.
\newblock {\em Theoretical and Computational Fluid Dynamics 2009 24:1},
  24(1):291--297, jul 2009.

\bibitem{Krueger2005}
P.~S. Krueger and M.~Gharib.
\newblock {Thrust augmentation and vortex ring evolution in a fully pulsed
  jet}.
\newblock {\em AIAA Journal}, 43(4):792--801, 2005.

\bibitem{Levy1927}
H.~Levy and A.~G. Forsdyke.
\newblock {The Stability of an Infinite System of Circular Vortices}.
\newblock {\em Proceedings of the Royal Society A: Mathematical, Physical and
  Engineering Sciences}, 114(768):594--604, apr 1927.

\bibitem{Linden2001}
P.~F. Linden and J.~S. Turner.
\newblock {The formation of ‘optimal' vortex rings, and the efficiency of
  propulsion devices}.
\newblock {\em Journal of Fluid Mechanics}, 427:61--72, jan 2001.

\bibitem{Meleshko2010}
V.~V. Meleshko.
\newblock {Coaxial axisymmetric vortex rings: 150 years after Helmholtz}.
\newblock {\em Theoretical and Computational Fluid Dynamics}, 24(1-4):403--431,
  mar 2010.

\bibitem{Mohseni2006}
K.~Mohseni.
\newblock {Pulsatile vortex generators for low-speed maneuvering of small
  underwater vehicles}.
\newblock {\em Ocean Engineering}, 33(16):2209--2223, nov 2006.

\bibitem{Norbury1973}
J.~Norbury.
\newblock {A family of steady vortex rings}.
\newblock {\em Journal of Fluid Mechanics}, 57(3):417--431, 1973.

\bibitem{Northrup1911}
E.~F. Northrup.
\newblock {An experimental study of vortex motions in liquids}.
\newblock {\em Journal of the Franklin Institute}, 172(3):211--226, 1911.

\bibitem{Oroskar2006}
S.~Oroskar, D.~R. Jackson, and D.~R. Wilton.
\newblock {Efficient computation of the 2D periodic Green's function using the
  Ewald method}.
\newblock {\em Journal of Computational Physics}, 219(2):899--911, dec 2006.

\bibitem{Saffman1993}
P.~G. Saffman.
\newblock {Axisymmetric Vortex Rings}.
\newblock In {\em Vortex Dynamics}, pages 192--200. Cambridge University Press,
  1993.

\bibitem{Sallet1975}
D.~W. Sallet.
\newblock {Impulsive motion of a circular disk which causes a vortex ring}.
\newblock {\em Physics of Fluids}, 18(1):109--111, 1975.

\bibitem{Siekmann1963}
J.~Siekmann.
\newblock {On a pulsating jet from the end of a tube, with application to the
  propulsion of certain aquatic animals}.
\newblock {\em Journal of Fluid Mechanics}, 15(3):399--418, 1963.

\bibitem{Strogatz2018}
S.~H. Strogatz.
\newblock {\em {Nonlinear Dynamics and Chaos}}.
\newblock CRC Press, may 2018.

\bibitem{Sullivan2008}
I.~S. Sullivan, J.~J. Niemela, R.~E. Hershberger, D.~Bolster, and R.~J.
  Donnelly.
\newblock {Dynamics of thin vortex rings}.
\newblock {\em Journal of Fluid Mechanics}, 609:319--347, 2008.

\bibitem{tait1867pm}
P. G. Tait.
\newblock {\em On integrals of the hydrodynamical equations, which express vortex-motion}.
\newblock {\em Philosophical Magazine}, 4:485-512, 1867.

\bibitem{Taylor1953}
G.~I. Taylor.
\newblock {Formation of a vortex ring by giving an impulse to a circular disk
  and then dissolving it away}.
\newblock {\em Journal of Applied Physics}, 24(1):104, 1953.

\bibitem{Thomson1883}
J.~J. Thomson.
\newblock {\em {A Treatise on the Motion of Vortex Rings: An Essay to which the
  Adams Prize was Adjudged in 1882, in the University of Cambridge}}.
\newblock Cambridge University Adams Prize Essay. Macmillan, 1883.

\bibitem{Thomson1867}
W.~Thomson.
\newblock {The translatory velocity of a circular vortex ring}.
\newblock {\em The London, Edinburgh, and Dublin Philosophical Magazine and
  Journal of Science}, 33:511--512, 1867.
  
\bibitem{Thomson1867a}
W.~Thomson.
\newblock {On vortex atoms}.
\newblock {\em The London, Edinburgh, and Dublin Philosophical Magazine and
  Journal of Science}, 34(227):15--24, jul 1867.

\bibitem{Tietjens1934}
O.~G. Tietjens and L.~Prandtl.
\newblock {\em {Fundamentals of hydro- and aeromechanics}}.
\newblock Dover Publications, 1934.

\bibitem{Vasilev1913}
N.~S. Vasilev.
\newblock {Reduction of the Equations of Motion of Coaxial Vortex Rings to
  Canonical Form}.
\newblock {\em Zap. Mat. Otd. Novoross. Obs. Est. ( Notes of the Mathematical
  Department of the Novorossiysk Society of Naturalists)}, 21:1--12, 1913.

\bibitem{Vasilev1916}
N.~S. Vasilev.
\newblock {On the motion of an infinite series of circular vortex rings of the
  same radius having a common axis}.
\newblock {\em Zap. Fiz.-Mat. Fak. Imp. Novoross. Univ (Notes of the Physics
  and Mathematics Department of the Imperial Novorossiysk University)},
  10:1--44, 1916.

\end{thebibliography}

\endpaper

\end{document}